\documentclass[twocolumn]{aastex631}

\usepackage{amsmath}
\usepackage{rotating}

\begin{document}

\title{A second radio flare from the tidal disruption event AT2020vwl: a delayed outflow ejection?}

\author[0000-0003-3441-8299]{A. J. Goodwin}
\affiliation{International Centre for Radio Astronomy Research – Curtin University, GPO Box U1987, Perth, WA 6845, Australia}
\author{A. Mummery}
\affiliation{Oxford Theoretical Physics, Beecroft Building, Clarendon Laboratory, Parks Road, Oxford, OX1 3PU, United Kingdom}
\author[0000-0003-1792-2338]{T. Laskar}
\affiliation{Department of Physics \& Astronomy, University of Utah, Salt Lake City, UT 84112, USA}
\author[0000-0002-8297-2473]{K. D. Alexander}
\affiliation{Department of Astronomy/Steward Observatory, 933 North Cherry Avenue, Rm. N204, Tucson, AZ 85721-0065, USA}
\author[0000-0001-6544-8007]{G. E. Anderson}
\affiliation{International Centre for Radio Astronomy Research – Curtin University, GPO Box U1987, Perth, WA 6845, Australia}
\author{M. Bietenholz}
\affiliation{SARAO/Hartebeesthoek Radio Observatory, PO Box 443, Krugersdorp 1740, South Africa}
\author{C. Bonnerot}
\affiliation{School of Physics and Astronomy, University of Birmingham, Birmingham B15 2TT}
\affiliation{Institute for Gravitational Wave Astronomy, University of Birmingham, Birmingham B15 2TT}
\author[0000-0003-0528-202X]{C. T. Christy}
\affiliation{Department of Astronomy/Steward Observatory, 933 North Cherry Avenue, Rm. N204, Tucson, AZ 85721-0065, USA}
\author{W. Golay}
\affiliation{Center for Astrophysics $|$ Harvard \& Smithsonian, 60 Garden St., Cambridge, MA 02138, USA}
\affiliation{Department of Astrophysics/IMAPP, Radboud University, P.O. Box 9010, 6500 GL, Nijmegen, The Netherlands}
\author{W. Lu}
\affiliation{Departments of Astronomy and Theoretical Astrophysics Center, UC Berkeley, Berkeley, CA 94720, USA}
\author{R. Margutti}
\affiliation{Department of Astronomy, University of California, Berkeley, CA 94720-3411, USA}
\affiliation{Department of Physics, University of California, 366 Physics North MC 7300, Berkeley, CA 94720, USA}
\author{J. C. A. Miller-Jones}
\affiliation{International Centre for Radio Astronomy Research – Curtin University, GPO Box U1987, Perth, WA 6845, Australia}
\author{E. Ramirez-Ruiz}
\affiliation{Department of Astronomy and Astrophysics, University of California, Santa Cruz, CA 95064, USA}
\author{R. Saxton}
\affiliation{Telespazio UK for ESA, European Space Astronomy Centre, Operations Department, 28691 Villanueva de la Canada, Spain}
\author{S. van Velzen}
\affiliation{Leiden Observatory, Leiden University, Postbus 9513, 2300 RA Leiden, The Netherlands}

\begin{abstract}

We present the discovery of a second radio flare from the tidal disruption event (TDE) AT2020vwl via long-term monitoring radio observations.
% Introduction
%The `gap' or problem
Late-time radio flares from TDEs are being discovered more commonly, with many TDEs showing radio emission 1000s of days after the stellar disruption, but 
the mechanism that powers these late-time flares is uncertain. 
Here we present radio spectral observations of the first and second radio flares observed from the TDE AT2020vwl. 
%The overall approach, key results, and conclusions
Through detailed radio spectral monitoring, we find evidence for two distinct outflow ejection episodes, or a period of renewed energy injection into the pre-existing outflow. We deduce that the second radio flare is powered by an outflow that is initially slower than the first flare, but carries more energy and accelerates over time. Through modelling the long-term optical and UV emission from the TDE as arising from an accretion disc, we infer that the second radio outflow launch or energy injection episode occurred approximately at the time of peak accretion rate. The fast decay of the second flare precludes environmental changes as an explanation, while the velocity of the outflow is at all times too low to be explained by an off-axis relativistic jet. 
%The advance over previous work, the implications
Future observations that search for any link between the accretion disc properties and late time radio flares from TDEs will aid in understanding what powers the radio outflows in TDEs, and confirm if multiple outflow ejections or energy injection episodes are common. 
\end{abstract}

\keywords{radio continuum: general; black hole physics; accretion, accretion disks}

\section{Introduction} \label{sec:intro}
A tidal disruption event (TDEs) occurs when a star in the center of a galaxy passes too close to the supermassive black hole and is destroyed \cite[e.g.,][]{Rees1988}. TDEs produce bright flares observed across the electromagnetic spectrum \citep[e.g.,][]{Auchettl2017,Alexander2020,vanvelzen2020}, providing a unique opportunity to witness the real-time evolution of accretion onto a previously dormant SMBH. During a TDE, approximately half of the stellar debris remains on bound orbits to eventually be accreted by the SMBH, while the other half is unbound \citep{Rees1988}, and can be ejected at significant velocities. 

In recent years, dozens of new TDEs have been discovered, primarily via transient emission in the optical, UV, or soft X-ray
\citep[e.g.,][]{vanvelzen2020}. In the optical and UV, TDEs are characterised by a 1-2\,mag rise in the optical flux of the nucleus of a quiescent galaxy (although TDEs can occur in active galaxies they are more difficult to classify), followed by a longer decay that approximately follows the theoretical fall-back rate onto a SMBH, $L\propto t^{-5/3}$ \citep[e.g.,][]{Phinney1989,Mockler2019,vanvelzen2021,Hammerstein2023,Yao2023}. 
Optical spectral observations of TDEs usually show transient broad H and/or He lines \citep[e.g.,][for a recent study]{Yao2023} and can show blueshifted emission lines interpreted as signatures of outflowing material \citep[e.g.,][]{Nicholl2020}. The optical/UV emission from TDEs is well-described by thermal blackbody emission with a temperature of $\sim10^4$\,K \citep[e.g.,][]{Gezari2009, Guillochon2014, vanvelzen2020}. Simulations of TDEs have shown that the optical/UV emission may be powered by an accretion disc that forms rapidly after the disruption \citep[e.g.,][]{Guillochon2014b,Shiokawa2015,Bonnerot2016,Hayasaki2016,Mummery2020}, debris collisions and shocks \citep[e.g.,][]{Dai2015,Lu2020}, or an envelope of debris that reprocesses higher energy emission from closer to the SMBH \citep[e.g.,][]{Loeb1997,Metzger2016,Roth2016,Metzger2022}. 

Long-term optical monitoring of TDEs has revealed a flattening and excess of optical emission present in approximately two thirds of optical lightcurves years post initial flare \citep{Mummery2024}. These late-time optical plateaus are well modelled by emission from an accretion disc with continued accretion onto the SMBH at significantly sub-Eddington rates \citep{Mummery2024}. Long-term optical monitoring of TDEs has also revealed a handful of events that show unusual late-time behaviour, including ones that rebrighten \citep[e.g.,][]{Yao2023,Somalwar2023}, interpreted to be due to a partial disruption of the star. 

X-ray emission from TDEs is less uniform, with some showing thermal X-ray emission that tends to decay within months of the optical flare \citep[e.g.,][]{Guolo2024}, although not all events produce detectable X-ray emission. Others show re-brightening or rapid variability in the X-rays \citep[e.g.,][]{Malyali2024}, which in some cases has also been interpreted as due to a partial disruption of the star \citep[e.g.,][]{Liu2023,Liu2024,Malyali2023b} or accretion disk changes \citep{Thomsen2022}. 

Radio observations of TDEs probe synchrotron emission from electrons that are accelerated in the shocks formed from outflows and jets. Outflows and jets from TDEs have been proposed to be launched by accretion-induced winds or magnetic field induced ejections from the disc \citep[e.g.,][]{vanvelzen2016}, debris collisions \citep[e.g.,][]{Goodwin2022}, or the unbound debris stream interacting with the circumnuclear medium \citep[e.g.,][]{Krolik2016}. However, radio detections of TDEs are uncommon, with $<30$ TDEs having published radio detections \citep[e.g.][]{Alexander2020}, and only a handful have been monitored continuously from the onset of the outflow \citep[e.g.,][]{Goodwin2022}. 
Very rarely ($\sim1\%$ of observed events), a TDE produces a highly-collimated, relativistic, radio-emitting jet \citep{Levan2011,Burrows2011,Zauderer2011,Bloom2011,Andreoni2022,Pasham2023}, whilst others present slower-moving, likely less-collimated radio-emitting outflows \citep[e.g.,][]{Alexander2016,Cendes2021,Stein2021,Christy2024,Goodwin2024}. 

TDEs show diverse radio properties and the origin of these differences is not well understood. Recent radio observations have complicated the theoretical interpretation of the radio emission from TDEs by illuminating that at least 40$\%$ show late-time ($\gtrsim 500$\,d) radio flaring, apparently not associated with optical or X-ray re-flares \citep{Cendes2023,Horesh2021,Anumarlapudi2024}. The source of these late time flares is currently being hotly debated by the community, with the leading scenarios being a delayed radio jet or renewed energy injection from accretion \citep[e.g.,][]{Cendes2023}, an off-axis jet coming into view \citep[e.g.,][]{Matsumoto2023}, or new interactions with inhomogenous clouds in the circumnuclear medium \citep[CNM;][]{Matsumoto2024,Zhuang2024}. To date, most TDEs that show late time ($\gtrsim 500$\,d) rising radio emission do not have comprehensive radio observations at early times, making it difficult to determine if the late time radio emission is linked to outflows produced at the time of the initial stellar disruption and if it is a re-brightening of a radio flare at earlier times. 

The TDE AT2020vwl (also Gaia20etp, ZTF20achpcvt) was discovered on 2020 October 10 by the Gaia Spacecraft as a $\sim1$\,mag optical flare at the centre of the galaxy SDSS J153037.80+265856.8/LEDA 1794348 at a distance of 147\,Mpc \citep[J2000 RA, Dec 15:30:37.800, +26:58:56.89][]{Hodgkin2020}. The event was also observed by the Zwicky-Transient Facility (ZTF) \citep{Yao2023} and classified by \citet{Hammerstein2021} as a TDE based on the optical spectral properties, bright UV-flux, and optical lightcurve. A prompt radio-emitting outflow was observed associated with the event, and attributed to an outflow likely produced during the circularisation of the stellar debris in the first tens of days post-disruption \citep{Goodwin2023}.

In this work, we present long-term radio observations of the TDE AT2020vwl spanning 97--1220\,d post-optical flare. Our observations reveal a second radio flare beginning approximately 400\,d after the optical flare began, in addition to the first radio flare reported by \citet{Goodwin2023}. These observations track the evolution of both the first and second radio flares, providing a unique opportunity in this work to constrain the mechanism of the late time rising radio emission in this source. 

The paper is laid out as follows: in Section \ref{sec:observations} we describe the new radio observations and data reduction processes, in Section \ref{sec:lightcurve_fitting} we present and model the radio lightcurve at different frequencies, in Section \ref{sec:outflowmodelling} we model the radio spectra and physical outflow properties, in Section \ref{sec:discmodelling} we present the disc model fits to the multiwavelength lightcurves of the event, in Section \ref{sec:discussion} we discuss the implications of the outflow and disc constraints, including possible explanations for the second radio flare, and finally in Section \ref{sec:conclusion} we provide a summary of the results.

\section{Observations}\label{sec:observations}
We observed AT2020vwl on multiple occasions with the Karl G. Jansky Very Large Array (VLA), upgraded Giant Metrewave Radio Telescope (uGMRT), and MeerKAT radio telescope between 2022 November 1 and 2024 March 26 at various frequencies in the range 0.6--16.6\,GHz. In addition \citet{Goodwin2023} reported 7 epochs of radio observations between 2021 Feb 3 and 2022 May 8 which we also include in our analysis. A summary of the new radio observations and measured flux densities of AT2020vwl is provided in the Appendix in Table \ref{tab:radio_obs}. 

\subsection{VLA}
We observed the coordinates of AT2020vwl on 5 occasions with the VLA  between 2022 November 1 and 2024 March 26 at 1-16.6\,GHz. All data were reduced using the same method as in \citet{Goodwin2023}, with calibration and imaging performed in the Common Astronomy Software Application \citep[CASA; version 5.6.3;][]{CASA2007,CASA2022} using standard procedures including the VLA calibration pipeline (version 5.6.3). Flux and bandpass calibration was done with 3C 286 in all observations. Observations were carried out at a central frequency of 1.5\,GHz (L-band) with 1\,GHz of bandwidth, 3\,GHz (S-band) with 2\,GHz of bandwidth, 6\,GHz (C-band) with 4\,GHz of bandwidth, 10\,GHz (X-band) with 4\,GHz of bandwidth, and 15\,GHz (Ku-band) with 6\,GHz of bandiwdth. 8-bit samplers were used for L- and S-band and 3-bit samplers were used for Ku-, X-, and C- bands. For phase calibration we used ICRF J151340.1+233835 for 2–18 GHz (Ku-, X-, C-, and S-band); and ICRF J160207.2+332653 for 1–2 GHz (L-band).

\subsection{GMRT}
We observed AT2020vwl with the upgraded Giant Metrewave Radio Telescope (uGMRT) on four occasions between 2023 Apr and 2023 Nov. The observations were taken in band 4, with a central frequency of 0.65 GHz and total bandwidth of 300 MHz; and band 5, with a central frequency of 1.26GHz and total bandwidth of 400 MHz. 
Each observing band was broken into 2048 spectral channels. Data reduction was carried out using the same procedures as in \citet{Goodwin2023}, including standard procedures in CASA (version 5.6.3). Flux and bandpass calibration was done with 3C286 and phase calibration with ICRF J160207.2+332653. The flux density of the target was again extracted in the image plane using imfit by fitting an elliptical Gaussian fixed to the synthesised beam. Unfortunately the band 5 observation taken in 2023 April suffered from issues with the flux calibration scan resulting in poor flux calibration of the data. We therefore do not include this observation in our analysis. 

\subsection{MeerKAT}
We observed AT2020vwl with the South African MeerKAT radio telescope on 2024-02-27 at a central frequency of 0.8\,GHz. We used the 4K (4096-channel) wideband continuum mode with an observed bandwidth from 544 to 1088 MHz, with a central frequency of 816 MHz. The data were reduced using the same procedures outlined in \citet{Goodwin2023} including the OxKAT scripts (Heywood 2020). We used  ICRF J133108.2+303032 (3C 286) and ICRF J193925.0-634245 to set the flux density scale and calibrate the bandpass and ICRF J160913.3+264129 (QSO B1607+268) as a secondary calibrator. 
The flux density of AT2020vwl was determined by fitting an elliptical Gaussian of the same dimensions as the restoring beam.

\subsection{Swift}
The Neil Gehrels Swift Observatory \citep[Swift;][]{Gehrels2004} observed AT2020vwl 28 times between 2020 Jan 07 and 2023 June 27. \citet{Goodwin2023} reported no X-ray detection of the event in the first 27 observations. We examined the 28th Swift-XRT observation taken on 2023 June 27 using the online Swift product builder \citep{Evans2009}. Again, there was no X-ray source detected at the position of AT2020vwl, with a 3$\sigma$ upper limit of on the 0.3--10 keV X-ray flux $F_X < 3.1 \times 10^{-13}$\,erg\,cm$^{-2}$\,s$^{-1}$, assuming a Galactic hydrogen column density of $N_H = 4.3 \times 10^{20}$\,cm$^{-2}$ \citep{Willingale2013} and photon index $\Gamma=1.5$ as appropriate for TDE X-ray emission \citep[e.g.,][]{Guolo2024,Auchettl2017}. %We report all Swift-XRT X-ray upper limits in Table \ref{tab:x-rayanduvobs}, extracted using the online Swift product builder.

\subsection{Interstellar scintillation}
Variability in the radio emission of a compact extragalactic radio source may be induced by interstellar scintillation \citep[ISS][]{Walker1998}. \citet{Goodwin2023} included an additional uncertainty on the observed radio flux densities at all frequencies to account for additional uncertainty in the radio flux density of the compact source. As discussed by \citet{Goodwin2023}, for the Galactic coordinates of AT2020vwl, the radio emission from a compact source will be in the strong, refractive regime until the source reaches an angular size of 134 microarcseconds ($\sim10^{17}$\,cm at the distance of AT2020vwl). At source sizes greater than this, we do not expect interstellar scintillation to cause significant variability in the radio measurements. Since the minimum radio source size is estimated to be largely $<10^{17}$\,cm in all new observations presented in this paper, we conservatively include an additional, frequency dependent uncertainty for the spectral modelling. As in \citet{Goodwin2023}, these additional uncertainties on the observed flux density range from 40$\%$ at 1.5\,GHz to 2$\%$ at 18\,GHz.     

\begin{figure}[ht!]
\includegraphics[width=\columnwidth]{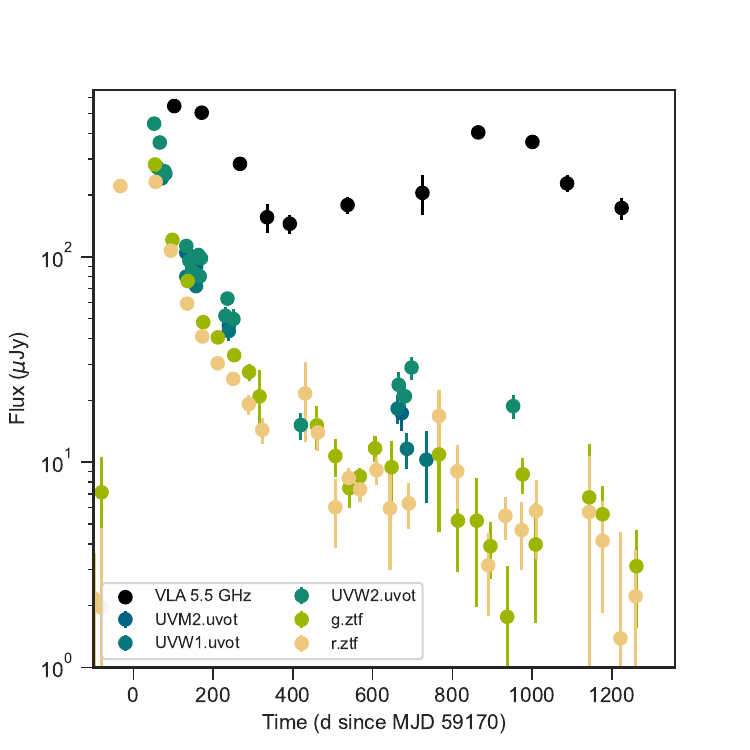}
\caption{The optical (ZTF g- and r-band), UV (Swift UVOT), and 5.5\,GHz radio (VLA) lightcurves of AT2020vwl. The optical and UV fluxes have been host-subtracted. The second radio flare does not appear to be correlated with any significant renewed optical or UV activity.
\label{fig:rad_lc}}
\end{figure}

\section{Lightcurve fitting}\label{sec:lightcurve_fitting}

The optical, UV, and 5.5\,GHz radio lightcurves of AT2020vwl are shown in Figure \ref{fig:rad_lc}. The optical flare began on approximately MJD 59130 \citep{Goodwin2023}, and radio observations did not begin until 142\,d later, at which time fading radio emission was detected. 

\begin{figure}
    \centering
    \includegraphics[width=\columnwidth]{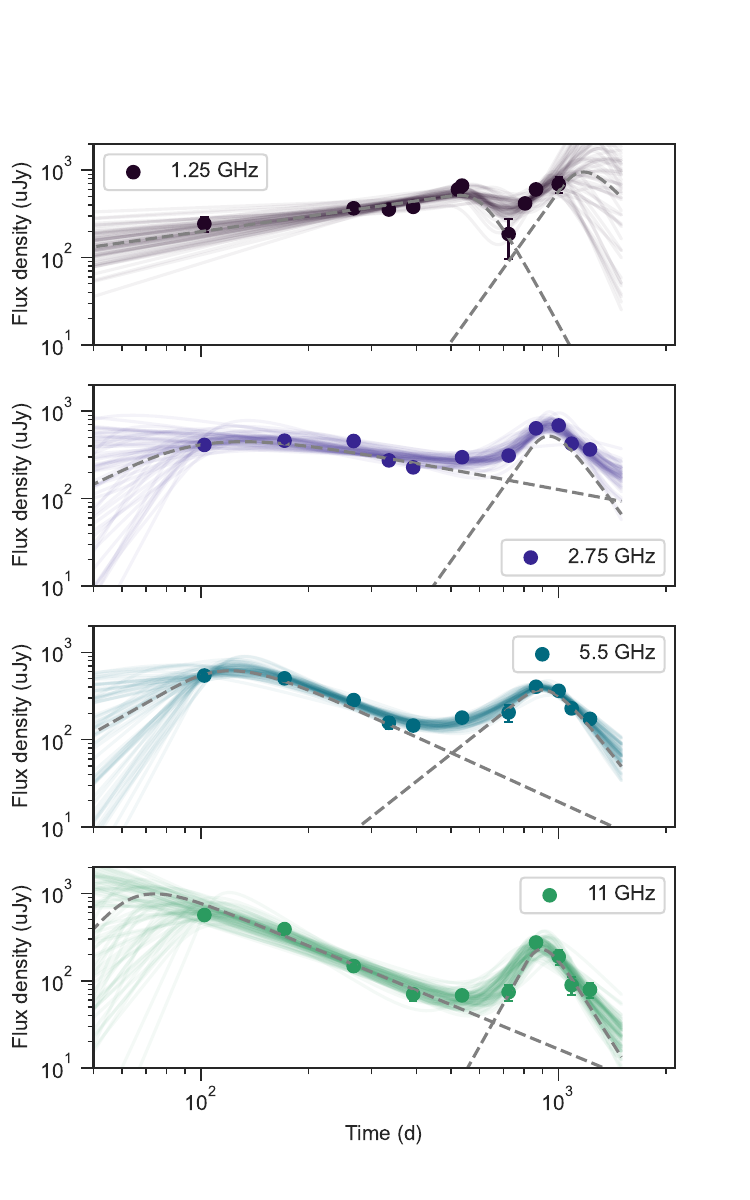}
    \caption{The 1.25, 2.75, 5.5, and 11\,GHz observed lightcurves of AT2020vwl (scatter points) and the two-component broken power-law fit at each frequency (solid lines). In each panel 100 random draws from the MCMC posterior distribution are plotted to indicate the uncertainty on the fit. The dashed grey lines show the individual power-law components for the first and second flares, demonstrating that the radio emission from the first power-law component is almost negligible during the second flare, except at the lowest observing frequency of 1.25\,GHz.}
    \label{fig:lcfits}
\end{figure}

In the first 400\,d post-optical flare, the radio emission from AT2020vwl decayed in a frequency-dependent manner with the decay occurring at higher frequencies first, as was reported by \citet{Goodwin2023}. However, between 400 and 600\,d post optical flare, our new radio observations reveal that the radio emission began rising at all observed frequencies (Figure \ref{fig:lcfits}). The radio emission rose quickly (over 100-200\,d), and began decaying at frequencies $>2$\,GHz between 800-900\,d post-optical flare. 

In order to constrain the rate of rise and decay of the radio emission at the different observed frequencies, we fit the individual frequency lightcurves with a broken power-law model. Following the \citet{Chevalier1998} prescription for modelling the time evolution of a self-absorbed synchrotron source, we assume the observed flux density at each frequency, $F_{\nu}$ is described by

\begin{equation}\label{eq:broken_plaw}
    F_{\nu} = 1.582 F_{\nu,tc} \left(\frac{t}{t_c}\right)^a \left[1 - \exp\left(-\left(\frac{t}{t_c}\right)^{-(a+b)}\right)\right]
\end{equation}

where $F_{\nu,tc}$ is the flux density at frequency $\nu$, $F_{\nu,tc}$ is the flux density at the break time, $t_c$ and $a$ and $b$ the powerlaw slope either side of the break time, where each observed frequency has different $t_c$, $a$, and $b$. In order to model the two radio flares, we assume two broken powerlaws such that 

\begin{equation}\label{eq:2broken_plaw}
    F_{\nu,\rm{total}} = F_{\nu,1} + F_{\nu,2}
\end{equation}

where $F_{\nu,1}$ and  $F_{\nu,2}$ (and associated $t_{c,1}$, $t_{c,2}$ etc.) are each given by Equation \ref{eq:broken_plaw}. 

We use a Markov Chain Monte Carlo (MCMC) approach to fit Eq \ref{eq:2broken_plaw} to the radio lightcurves at the observed frequencies of 1.25, 2.75, 5.5, and 11\,GHz. We use the Python implementation of MCMC, \texttt{emcee} \citep{emcee2013} and flat prior distributions on all modelled parameters, where we allow the parameters to vary between:
$0.1 < \rm{log}\,F_{\nu,tc}\,(\rm{\mu Jy}) < 6$; $0 < a < 10$; $0 < b<10$; $400 < t_c\,(\rm{d}) < 1500$. We ran each chain with 200 walkers for 3000 steps, conservatively discarding the first 2000 steps for burn-in. We note that for the lightcurve fitting we do not subtract a host component from the observations as we do for the spectral fitting in Section \ref{sec:specfits}, because the host component is constant at each frequency, very small compared to the observed flux densities, and would introduce an additional systematic error as it cannot be well-constrained until the transient emission no longer dominates the observed radio emission. 

Using this approach, we constrain the peak times of each of the two flares at each frequency as well as the power-law indices of the rising and decaying portions of the lightcurves. We note that the rise of the first radio flare is not as well constrained as the second flare because the radio observations did not begin until 142\,d post-optical flare. 
The modelled lightcurve parameters are listed in Table \ref{tab:lc_fits} and the corresponding best-fit lightcurves are plotted in Figure \ref{fig:lcfits}. 

\begin{table*}[]
    \centering
    \caption{Fitted lightcurve parameters using the two-component broken power-law model given in Equation \ref{eq:2broken_plaw}. The uncertainties correspond to the 67$\%$ confidence intervals.}
    % \begin{tabular}{p{1cm}p{1.5cm}p{1.5cm}p{1.5cm}p{1.5cm}p{1.5cm}p{1.5cm}p{1.5cm}p{1.5cm}}
     \begin{tabular}{lllllllll}
    \hline
    Freq. (GHz) & $F_{\nu,1}$\,($\rm{\mu}$Jy) & $t_{c,1}$ (d) & $a_1$ & $b_1$ & $F_{\nu,2}$\,($\rm{\mu}$Jy) & $t_{c,2}$ (d) & $a_2$ & $b_2$\\
    \hline
    \hline
1.25 & 368$^{+68}_{-54}$ & 628$^{+63}_{-41}$     & 0.6$^{+0.2}_{-0.2}$ & 7.6$^{+1.8}_{-3.9}$    & $>$955 & $>$1183 & 5.7$^{+2.9*}_{-2.2}$ & 4.6$^{+3.7*}_{-3.2}$ \\ 
2.75 & 433$^{+164}_{-106}$ & 120$^{+158}_{-68}$     & 1.6$^{+4.4}_{-1.3}$ & 0.8$^{+1.3}_{-0.4}$    & 529$^{+113}_{-122}$ & 958$^{+101}_{-68}$    & 5.7$^{+2.9*}_{-3.2}$ & 5.4$^{+2.4*}_{-2.3}$\\ 
5.5 & 606$^{+131}_{-179}$ & 127$^{+69}_{-22}$     & 2.1$^{+4.0}_{-2.0}$ & 1.9$^{+0.9}_{-0.5}$    & 351$^{+30}_{-42}$ & 961$^{+45}_{-52}$    & 3.1$^{+1.4}_{-0.8}$ & 5.4$^{+1.5}_{-1.2}$\\ 
11 & 890$^{+1767}_{-338}$ & 75$^{+39}_{-41}$     & 4.0$^{+3.8}_{-3.0}$ & 1.7$^{+0.5}_{-0.3}$    & 225$^{+52}_{-48}$ & 907$^{+45}_{-48}$    & 7.2$^{+2.0}_{-2.8}$ & 6.6$^{+1.9}_{-2.0}$\\   
    \hline
    \end{tabular}
    
\footnotesize{$^*$Parameter is unconstrained by the MCMC fit and consistent with returning the prior range.} 
    \label{tab:lc_fits}
\end{table*}

The lightcurve fits indicate significantly different properties for the first and second radio flares. Both radio flares peaked earliest at higher frequencies, with a difference of $\approx$50\,d between the 11\,GHz peak time and the 2.75\,GHz peak time for both the first and second flares. The first flare rose with a temporal power-law index between 1--4.4 across all frequencies, whilst the second flare rose with a temporal power-law index between 3.5--7.5, across all frequencies. The first flare decayed with a temporal power-law index between 0.5--3 whilst the second flare decayed with a temporal power-law index between 4--6.5. From these lightcurve fits it is clear that the second flare rose and decayed significantly more quickly at all frequencies than the first flare, but the delay between peak times at the observed frequencies was similar. 

\section{Spectral fitting and Outflow Modelling}\label{sec:outflowmodelling}
% General format:
% [link to previous paragraph? To [aim], we [method].
% We found [key feature of results (fig/table ref). [Key feature of results (fig/table ref)]. [Key feature of results (fig/table ref)]. This suggests [conclusion from these results].
% start the next paragraph with an aim that builds on the take-home-message, thus gradually building up the story of the paper.
\subsection{Spectral fits}\label{sec:specfits}
Next, we examined the radio spectra at each epoch of observations in order to constrain the evolution of the physical properties of the synchrotron-emitting region. We fit 11 epochs of observed radio spectra using the same approach outlined in \citet{Goodwin2023} including subtracting a host component then fitting the \citet{Granot2002} synchrotron spectral model. The exact equations used to fit the synchrotron flux density and constrain the synchrotron peak flux density, $F_{\rm{p}}$, minimum frequency, $\nu_{\rm{m}}$, peak frequency, $\nu_{\rm{p}}$, and synchrotron energy index, $p$, are given in Equations 2--4 of \citet{Goodwin2023}. We assume that the synchrotron emission is in the regime where $\nu_m < \nu_a < \nu_c$, where $\nu_c$ is the synchrotron cooling frequency. We use the same MCMC fitting procedure using the Python implementation of MCMC, \texttt{emcee} \citep{emcee2013}. Unlike in \citet{Goodwin2023}, we do not fit for both $\nu_m$ and $\nu_a$ for the new epochs presented in this work (2022-11-01 and onwards) as $\nu_m$ has likely shifted below the observed frequencies. Instead we fit for a single spectral break which we associate with $\nu_a$. The equation used to describe the synchrotron flux is given by \citet{Granot2002}, i.e.

\begin{equation}
    \label{eq:Fnua}
    \begin{aligned}
        F_{\nu, \mathrm{synch}} = F_{\nu,\mathrm{ext}} \left[
        \left(\frac{\nu}{\nu_{\rm a}}\right)^{-s\beta_1} +  \left(
        \frac{\nu}{\nu_{\rm a}}\right)^{-s\beta_2)
        }\right]^{-1/s}
        \end{aligned}
    \end{equation}
    where $\nu$ is the frequency, $F_{\nu,\mathrm{ext}}$ is the normalisation, $s = 1.25-0.18p$, $\beta_1 = \frac{5}{2}$, $\beta_2 = \frac{1-p}{2}$. 

\begin{figure}
    \centering
    \includegraphics[width=\columnwidth]{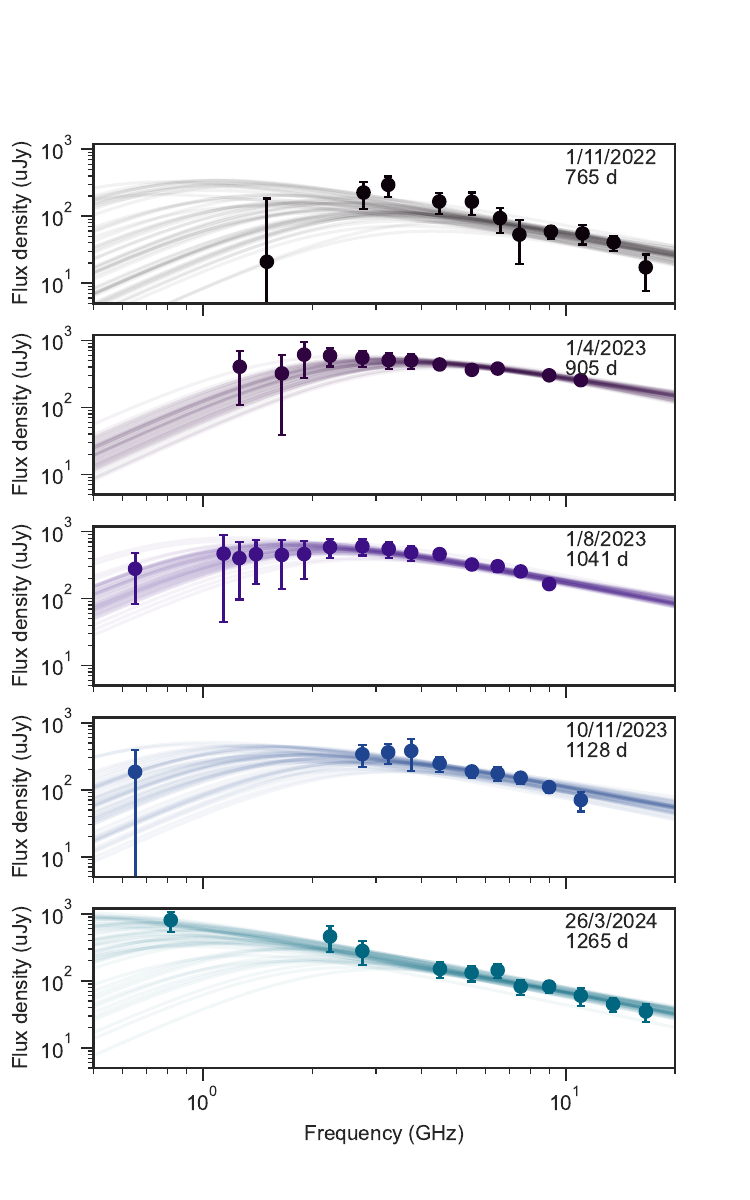}
    \caption{The observed (points) and modelled (lines) synchrotron spectra for AT2020vwl. Each panel is labelled with the UTC date of the observation and 100 random samples from the MCMC posterior are plotted to indicate the uncertainty in the fits.}
    \label{fig:specfits}
\end{figure}

We use the same prior distributions as in \citet{Goodwin2023} on all modelled parameters, except notably that we decreased the lower bound on the peak frequency to be 0.1\,GHz instead of 1\,GHz, as the peak frequency in the latest observations is significantly lower than in the earlier epochs.
The best-fit synchrotron spectra for each epoch are plotted in Figure \ref{fig:specfits} and the constrained parameters are listed in Table \ref{tab:specfits}. 

\begin{figure}
    \centering
    \includegraphics[width=\columnwidth]{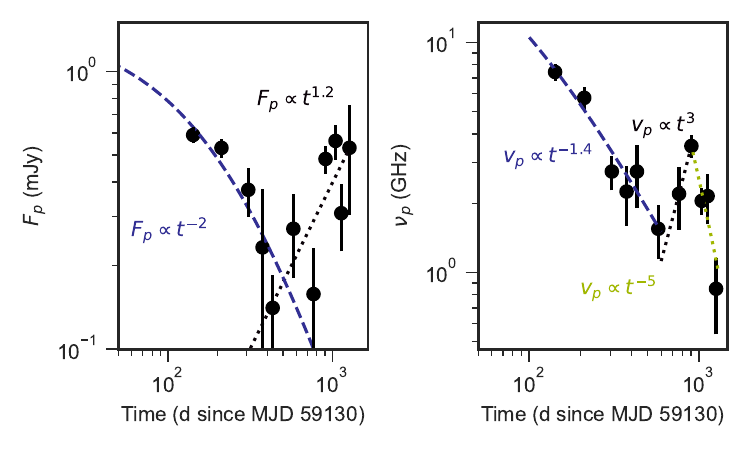}
    \caption{The modelled peak flux density (left panel) and peak frequency (right panel) of the synchrotron spectrum of AT2020vwl at each observed radio epoch. The peak flux density and frequency were initially decreasing during the first flare, but began increasing at the onset of the second flare (between 400-500\,d). The dotted and dashed lines show power-law fits to the rises and decays of the two parameters.}
    \label{fig:peakflux}
\end{figure}

The peak flux density and associated peak frequency of the modelled synchrotron spectra are plotted in Figure \ref{fig:peakflux}. The peak flux density was initially decreasing following $F_p\propto t^{-2}$, consistent with expectations for 
%as expected for 
% a synchrotron self-absorbed source 
synchrotron emission from a non-relativistic shock 
%evolving 
propagating 
in an environment with steep density stratification
(Appendix~\ref{sec:synappendix}), 
until between 400-500\,d after the onset of the optical flare, at which time it began increasing following $F_p\propto t^{1}$. The frequency of the peak was initially decreasing, consistent with an expanding synchrotron-emitting region, until $\approx$500\,d post-optical flare, at which time it increased for $\approx300$\,d, before beginning to decrease again. 

\subsection{Outflow model}
In order to constrain the evolution of the physical outflow properties, we model the synchrotron emission as a blastwave that accelerates ambient electrons into a power-law distribution, $N\propto \gamma^{-p}$, where $p$ is the synchrotron energy index, and $\gamma$ is the electron Lorentz factor. We use the same method outlined in \citet{Goodwin2023}, using the \citet{BarniolDuran2013} formalism and Equations 4--13 in \citet{Goodwin2022} to estimate key physical quantities such as the outflow radius ($R$), outflow energy ($E$), magnetic field strength ($B$), mass of the emitting region ($m_{\rm{ej}}$), ambient electron density ($n_e$), and the velocity of the ejecta ($\beta$). We first estimate the equipartition radius ($R_{\rm{eq}}$) and equipartition energy ($E_{\rm{eq}}$), then apply a correction to account for any deviation from equipartition. In this work we assume the fraction of the total energy in the electrons is 10$\%$ and the fraction of the total energy in the magnetic field is 2$\%$, but note that changes in the values of these fractions can cause significant changes in the derived physical outflow quantities, as described in detail by \citet{Goodwin2023}. However, the overall trends in the evolution of the physical outflow quantities remain the same for different assumptions about the deviation from equipartition as long as these deviations are time-independent. We provide outflow properties for two different geometries: a spherical emitting region and a conical emitting region corresponding to a mildly collimated jet with a half-opening angle of 30\,deg. The exact geometric factors that describe the geometry are the same as in \citet{Goodwin2023}.  

\begin{figure*}
    \centering
    \includegraphics[width=2\columnwidth]{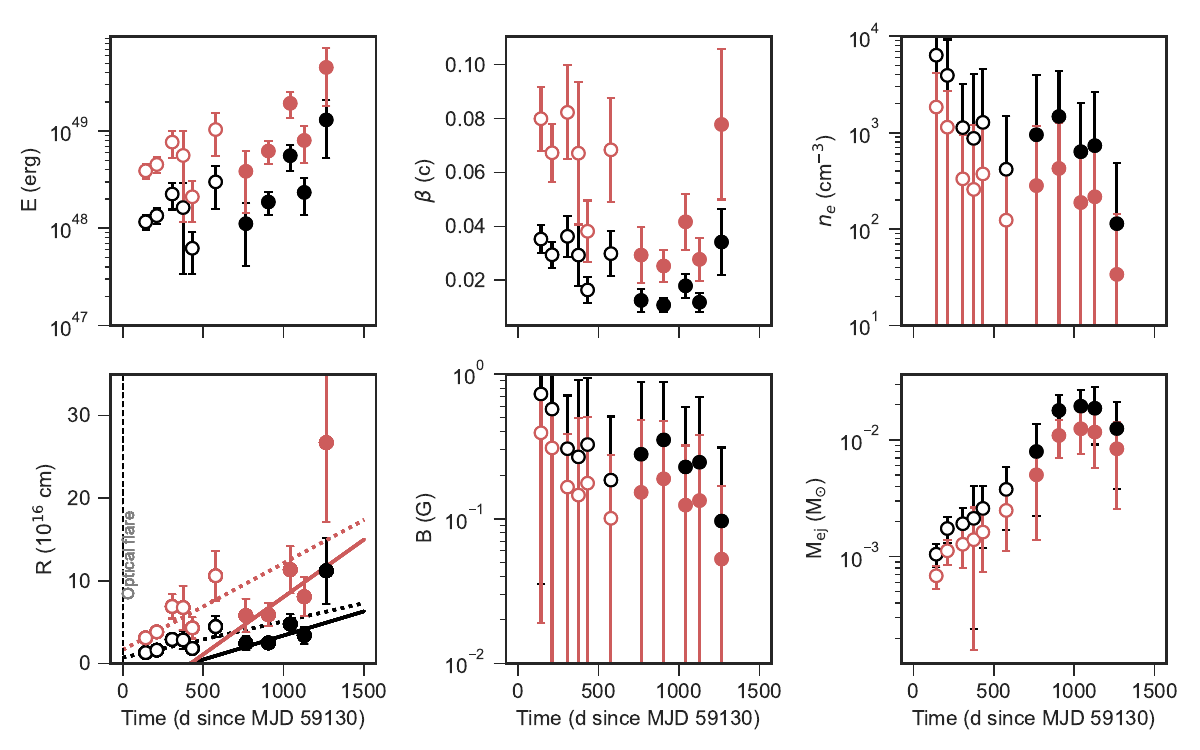}
    \caption{The outflow properties of AT2020vwl. Black points indicate outflow properties for a spherical geometry and red points for a conical geometry. Open circles indicate data published in \citet{Goodwin2023}, corresponding to the first radio flare. Closed circles indicate new data presented in this work, corresponding to the second radio flare. In the bottom left panel the dashed lines indicate linear fits to the radius for the first flare, and the solid lines indicate linear fits to the radius for the second flare.}
    \label{fig:outflowprops}
\end{figure*}

The outflow properties we infer are plotted in Figure \ref{fig:outflowprops} and listed in the Appendix in Table \ref{tab:specfits}. From the outflow model, we infer the following:
\begin{itemize}
    \item It appears that the energy in the outflow was initially constant, until the second flare began, at which time the energy began increasing. 

\item The velocity of the outflow was initially constant, until approximately the onset of the rise of the second flare, at which time it decreased by a factor of $\sim2$ and remained constant until the final epoch where an increase in the velocity is observed. 

\item The ambient density, whilst less well-constrained, initially was decreasing rather steeply with radius, until the second flare at which time it plateaued for a period before decreasing in the final epoch. 

\item The apparent radius of the outflow was initially increasing, and shows a drop or plateau at the time of the second flare, before continuing to increase again. 

\item The magnetic field strength, whilst poorly constrained, is consistent with being constant or decreasing across the course of the observations. 

\item The mass in the emitting-region increased until the peak of the second flare, after which time it remained approximately constant or decreased slightly.  
\end{itemize}

In Figure \ref{fig:outflowprops} it is evident that most of the constrained outflow parameters show a significant change from the initial trends at the time that the second radio flare began dominating the radio emission. This could be evidence of two distinct synchrotron-emitting regions that dominate the radio emission at different epochs. 
A drop in radius of the outflow would imply the outflow either had to move inwards, or the radio emission is dominated by a second, more compact emitting region at this epoch. Noting the large uncertainty on the estimated radius, if we assume two synchrotron-emitting regions, we can fit the radius over time with a linear fit for the first flare (up to 720\,d post-optical flare) and the second flare (after 720\,d post-optical flare).  We find for two independent synchrotron-emitting sources, the first was launched approximately at the time of optical discovery, whilst the second was launched between 160--700\,d after the optical discovery, where the range is determined due to differences in assumed geometry of the outflow, accounting for uncertainties on constraints from both geometries, and fitting including or excluding the final data point which is significantly higher than the rest of the data points. We note that these linear fits to the radius are overall poor fits to the data when the final data point is included, pointing towards variable velocity of the outflow. As an additional caveat we also note that excluding the data point at 765\,d, which may be affected by both emitting regions in the two-outflow scenario, results in an increase in the outflow launch time. 
In an alternate scenario, a plateau in the radius may be attributed to the initial deceleration of the outflow. Until, at the time of the second flare, renewed energy injection into the outflow causes the velocity to increase and the radius expands at a faster rate. The significantly increasing energy carried by the outflow evident in Figure \ref{fig:outflowprops} beginning at the time of the second flare is consistent with this scenario. 
Both of these scenarios require delayed energy injection hundreds of days after the TDE began.

\section{disc model}\label{sec:discmodelling}

\begin{figure*}
    \centering
    \includegraphics[width=\columnwidth]{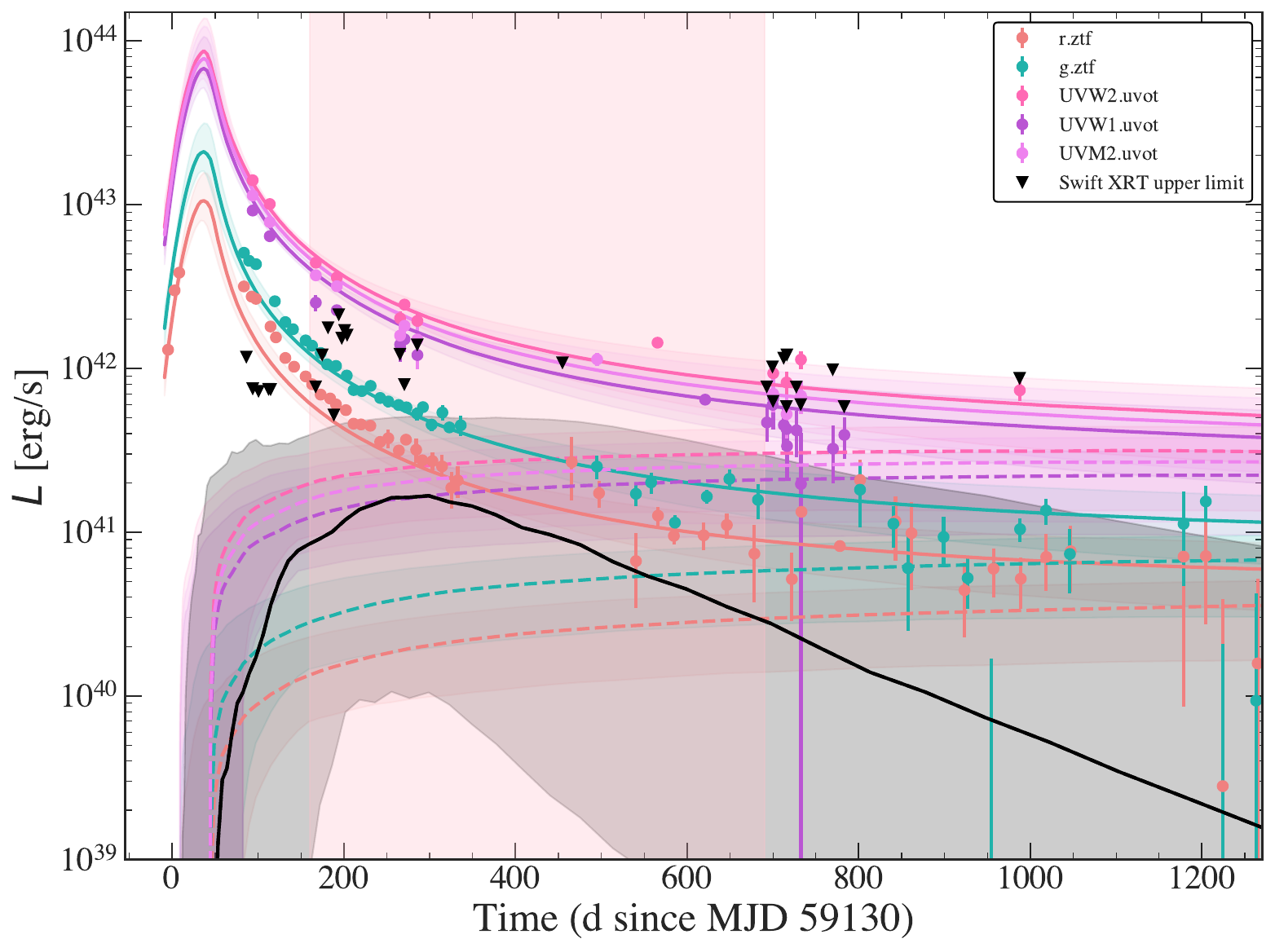}
    \includegraphics[width=\columnwidth]{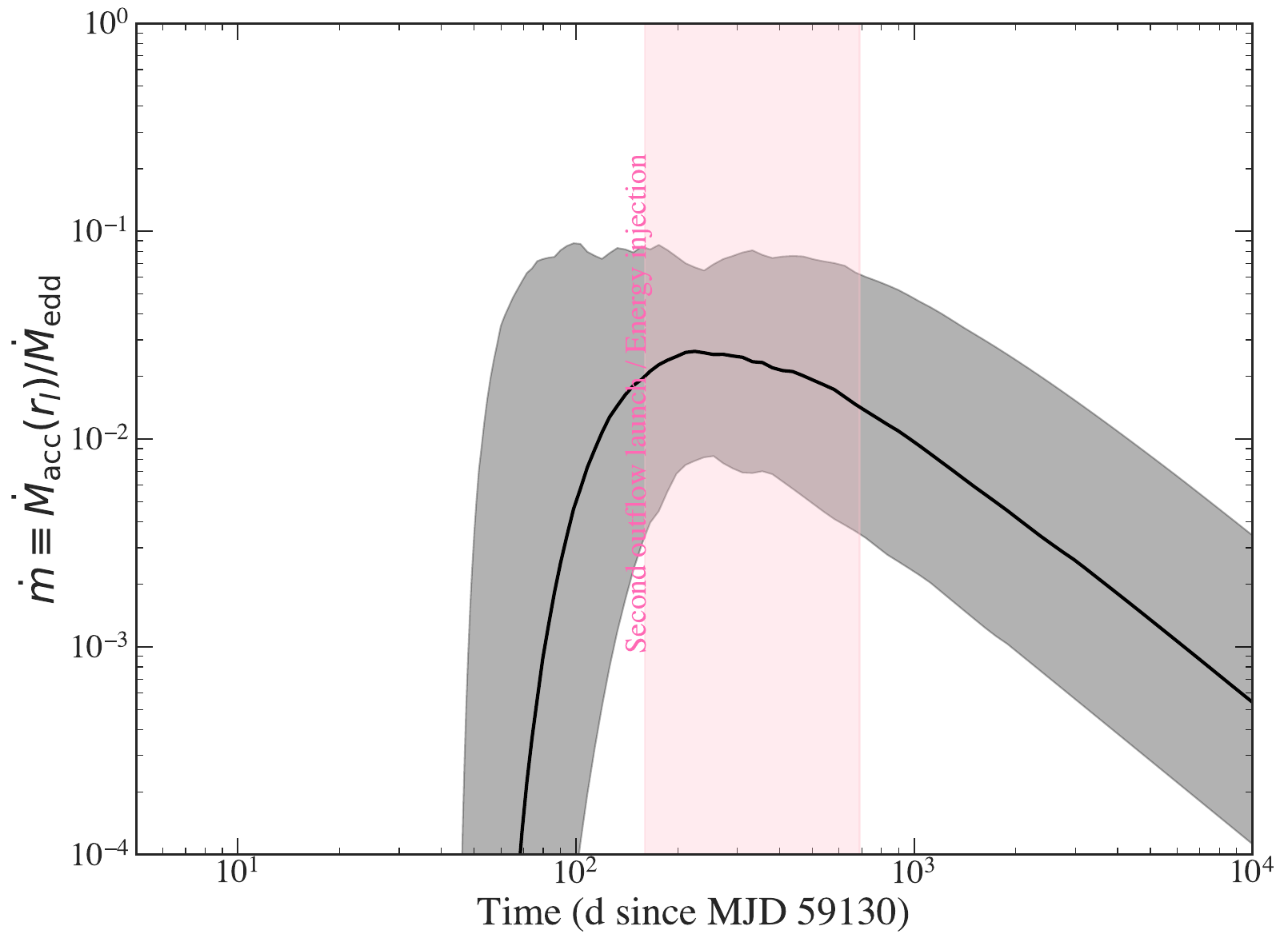}
    \caption{\textit{Left panel:} Model fits to the observed optical (red and green), UV (pink), and X-ray (black) lightcurves of AT2020vwl. Solid lines indicate the best fit model (Gaussian rise + power law decay + disc) with the shaded regions indicating the 68$\%$ credible intervals, and the dashed coloured lines show the contributions to the total luminosity from the disc model alone. Inverted triangles indicate 3$\sigma$ upper limits. \textit{Right panel:} The inferred time evolution of the accretion rate across the ISCO from the model posteriors (solid line) and the 68$\%$ credible interval (grey shading). The pink shaded region shows the approximate launch time of the second outflow, or the period at which energy injection is required, constrained via the synchrotron radius/energy measurements in Section \ref{sec:outflowmodelling}. The energy injection time is approximately coincident with the peak accretion rate reached in the disc model.}
    \label{fig:disc}
\end{figure*}

In common with most TDEs, the optical and UV lightcurves of AT2020vwl show a plateau at late times (Fig. \ref{fig:disc}), consistent with direct emission from an accretion flow. In this section we model this evolving accretion disc using the {\tt FitTeD} code developed by \cite{mummery2024fitted}. The {\tt FitTeD} code uses the \cite{Mummery23} solutions of the relativistic disc evolution equations \citep{Balbus17} to compute multi-band light curves, which we fit to the AT2020vwl optical/UV and X-ray data. We take the optical/UV data from the public {\tt manyTDE} repository \citep{Mummery2024}.  At optical/UV frequencies we include both a phenomenological power-law decay model, and a Gaussian rise model, to constrain the early (non-disc) emission. We also include the Swift X-ray upper limits.   

The key free parameters of the disc model are: the black hole mass $M_\bullet$ and (dimensionless) spin $a_\bullet$, the initial disc mass $M_{\rm disc}$, viscous timescale $t_{\rm visc}$, and disc formation radius $r_0$ and time $t_0$. The axis of the disc is inclined by an angle $i$
to the line of sight, also a fitting parameter. These parameters are inferred using standard MCMC techniques \citep{emcee2013}. The Markov chain converged and we find a best-fitting  black hole mass of $\log_{10} M_\bullet/M_\odot = 6.55^{+0.3}_{-0.5}$, broadly consistent with galactic scaling relationships: $\log_{10} M_\bullet/M_\odot = 5.6\pm 0.4$ \citep[scaling with galaxy velocity dispersion][]{Yao2023} and  $\log_{10} M_\bullet/M_\odot = 6.5 \pm 0.8$ \citep[scaling with galaxy mass][]{Mummery2024}. The constrained disc parameters are listed in Table \ref{tab:discmodel}. Note that the black hole spin parameter and the disc-observer inclination are poorly constrained owing to the lack of X-ray detections. This is physical, the spin and inclination more strongly affect observations probing the inner disc (X-ray) than outer disc (optical/UV), and so these parameters can not be well constrained for a source such as AT2020vwl which only has X-ray non-detections.

\begin{table}[]
    \caption{Modelled disc parameters for AT2020vwl. The uncertainties correspond to the 67$\%$ confidence intervals.}
    \centering
    \begin{tabular}{ll}
    \hline
         $\log_{10} M_\bullet/M_\odot$& $6.55^{+0.3}_{-0.5}$ \\
         $a_\bullet$ &  $-0.17^{+0.4*}_{-0.5}$\\
         $M_{\rm{disc}}$ ($M_\odot$) & 0.014$^{+0.022}_{-0.008}$ \\
         $r_0$ ($GM_\bullet/c^2$) & $40^{+50}_{-20}$ \\
         $t_{\rm{visc}}$ (d) &$833^{+122}_{-300}$ \\
         $t_0$ (d) & $6.63^{+82}_{-62}$ \\
         $i$ (deg) & $55^{+26*}_{-36}$ \\
         \hline
    \end{tabular}
    
\footnotesize{$^*$Parameter is unconstrained by the MCMC fit and consistent with returning the prior range.}
    \label{tab:discmodel}
\end{table}

In the left panel of Figure \ref{fig:disc} the model luminosity evolution of the accretion disc in the optical, UV, and X-ray observing bands is plotted along with the observed luminosities and X-ray upper limits (left panel). In the right panel we also show the inferred time evolution of the accretion rate across the inner-most stable circular orbit (ISCO) from the model posteriors. AT2020vwl was never detected in X-rays, and the optical/UV emission at late times is near time-independent, and so the late-time ISCO accretion rate inferred by the model is poorly constrained. The X-ray upper limits do however provide a strong constraint on the peak of the accretion rate reached in the AT2020vwl system, under the assumption that the X-rays are not obscured. Interestingly, we find the plausible launching time of the second radio outflow ($\Delta t \simeq 160-700$ days post peak), or renewed energy injection, approximately coincides with the time that the peak accretion rate was reached in the system. 

\section{Discussion}\label{sec:discussion}

\subsection{Did the outflow encounter a denser region of the CNM?}

It has been proposed that radio re-flaring events may be explained by a pre-existing outflow encountering a denser CNM that can re-accelerate the population of emitting electrons and produce a brightening in the radio emission. Denser clumps in the CNM may be due to an inhomogenous CNM \cite[e.g.,][]{Goodwin2022}, an increase in ambient density at the Bondi radius of the SMBH \citep[e.g.,][]{Matsumoto2024}, or one or more molecular clouds within the central region of the galaxy \citep[e.g.,][]{Zhuang2024}. 

In other types of synchrotron-emitting transients such as supernovae and gamma ray bursts (GRBs) from supernovae, radio lightcurves occasionally reveal re-flares or bumps \citep{Stroh2021}. These re-flares or bumps may be attributed to changes in the circumstellar environment \citep[e.g., SN 2007bg; SN2014C;][]{Salas2013,Anderson2017}, or renewed energy injection episodes \citep[e.g., GRB 210726A;][]{Schroeder2024}. Whilst the ejection mechanism and environment for supernovae outflows are significantly different from those of TDEs, the underlying physics of the synchrotron emission processes remain the same. In the case of changes in circumstellar environments of supernovae, transitions to denser environments can cause an increase in the radio flux density as the outflow experiences a stronger shock \citep{Salas2013}, and can result in spectral changes of the radio emission.

\citet{Matsumoto2024} proposed that late time radio flares observed in many TDEs may be explained by the outflow ejected at the time of the stellar disruption encountering the Bondi radius of the SMBH, at which time the CNM density gradient is expected to transition from being relatively steep to approximately flat. 
\citet{Matsumoto2024} describe, that for a CNM density profile of $n\propto R^{-k}$, the optically thin synchrotron luminosity, $\nu L_{\nu,\rm{thin}}$, evolves with time as

\begin{equation}\label{eq:Lthin}
    (\nu L_{\nu})_{\rm{thin}} \propto t^{\frac{12 -k(p+5)}{4}}.
\end{equation}

and the optically thick synchrotron luminosity evolves with time as

\begin{equation}\label{eq:Lthick}
    (\nu L_{\nu})_{\rm{thick}} \propto t^{\frac{k+8}{4}}.
\end{equation}

Taking $p\approx3$ for AT2020vwl (Table \ref{tab:specfits}), we therefore expect the optically thin radio emission (at the beginning of the rise of the second flare the radio emission from AT2020vwl was optically thin at frequencies $>1$\,GHz) to evolve as $(\nu L_{\nu})_{\rm{thin}} \propto t^{-1}$ for k=2 or $(\nu L_{\nu})_{\rm{thin}} \propto t^{-2}$ for $k=2.5$. At the Bondi radius, \citet{Matsumoto2024} propose the density gradient may change to $k=0$, which would result in the optically thin radio emission transitioning to evolve as $(\nu L_{\nu})_{\rm{thin}} \propto t^{3}$. Indeed, as found by \citet{Goodwin2023}, the initial decay of the radio lightcurves at all frequencies is well-described by an optically thin synchrotron emitting source with $k=2$ to 2.5 and $p=3$ (Table \ref{tab:lc_fits}). However, the second flare was observed to rise with $(\nu L_{\nu})_{\rm{thin}} \propto t^{3.5}$ to $t^{7.5}$, significantly steeper than the expected $t^3$ for a CNM transitioning from steep to flat at the Bondi radius. In order to explain the steeply rising gradient of the second radio flare observed, the CNM gradient would need to transition from $k\approx2$ to 2.5 to $k\approx-0.5$ to $-2.5$ (i.e. an increase in the density with radius from the SMBH). In addition, if the second radio decay were due to CNM interactions, the CNM gradient would need to transition from positive to extremely steeply decaying ($k\approx3$ to $5$) to explain the rapid decay of the radio emission after the peak of the second flare. The steep rising lightcurves and unlikely values of $k$ are therefore not consistent with the outflow encountering a transition in the CNM from steep to flat at the Bondi radius of the SMBH.

The sharp increase in the radio emission from AT2020vwl would require an extreme change in the density of the circumnuclear medium. \citet{Zhuang2024} suggested that the interaction of TDE outflows encountering a molecular cloud in the central region of the galaxy may explain some late time radio flares observed from TDEs. However, the increasing energy in the outflow observed during the second flare (Fig. \ref{fig:outflowprops}) is not explained by this scenario. The short duration of the second flare would require a small, dense molecular cloud interaction, but the drastic increase in energy observed is not compatible with this scenario. 
%Additionally,the increase in the peak frequency of the synchrotron spectrum that was observed implies a  tentative decrease in the minimum size of the radio-emitting region. If the second radio flare was triggered by the outflow encountering a molecular cloud, the radio-emitting region size should not decrease. 
We therefore find it unlikely that the re-brightening was triggered by a change in the density profile, or a localized densification, in the environment surrounding the outflow. 

\subsection{An off-axis relativistic jet?}

A late time rise in the radio emission may be caused by an off-axis relativistic jet launched at or near the time of the stellar disruption. An off-axis jet naturally produces a steep rise in the observed radio lightcurve as the jet decelerates and spreads laterally, with a larger fraction of the jet coming into the observer's line of sight at late times \citep{Matsumoto2023}. In the generalised equipartition model of \citet{Matsumoto2023}, the initially off-axis relativistic jet transitions to an on-axis Newtonian outflow at the time that the jet decelerates to sub-relativistic velocities. The equipartition Newtonian velocity at this time must be $\beta_{\rm{eq,N}}>0.23$ as for $\beta_{\rm{eq,N}}$ smaller than this the minimal energy for on- and off-axis solutions have discrete branches implying that extraordinarily high $\Gamma$ values and an extremely off-axis jet would be required to explain the radio luminosity. \citet{Matsumoto2023} define

\begin{equation}
\label{eq:betaeqN}
\begin{split}
    \beta_{\rm{Eq,N}} \approx 0.73 \left[ \frac{(F_p/\rm{mJy})^{8/17} (d_{\rm{L}}/10^{28}\rm{cm})^{16/17} \eta^{35/51}}{(\nu_p/10\rm{GHz}) (1+z)^{8/17}}\right]\\
    \times  \left(\frac{t}{100\,\rm{d}}\right)^{-1}
    f_A^{-7/17} f_V^{-1/17}
\end{split}
\end{equation}

where $d_{\rm{L}}$ is the luminosity distance, and $f_A$ and $f_V$ are the geometric factors that describe the geometry of the outflow. 

The peak flux density of the radio emission from AT2020vwl has been rising since the second flare began, with the final epoch at $t=1220$\,d having $F_p=0.6\pm0.4$\,mJy with $\nu_p=0.8\pm0.4$\,GHz. These values correspond to $\beta_{\rm{Eq,N}}=0.03$ (assuming $f_A=f_V=1$, as appropriate for a collimated jet), significantly lower than the required $\beta_{\rm{Eq,N}}=0.23$ for an outflow at the transition from relativistic to Newtonian. We therefore find it unlikely that the second radio flare is due to an off-axis relativistic jet, as the observed outflow velocity at the time of the flare is too low for this scenario. If the peak flux were to continue to rise for an extended time period, the outflow may reach the critical transition velocity of $\beta_{\rm{eq,N}}=0.23$. Although, given the observed flux density at frequencies $>2$\,GHz has already begun decreasing it is unlikely the peak flux density will continue to rise for sufficient time for this transition to occur.

\subsection{Renewed energy injection into the outflow?}

Increasing radio luminosity from a synchrotron-emitting outflow may be triggered by renewed energy injection into the outflow, or the launching of a new outflow. For example, there are a handful of radio flares observed in GRB afterglows that have been attributed to energy injection \citep[e.g.,][]{Schroeder2024}, and in X-ray binaries accretion disc state transitions are usually associated with radio jet launching \citep[e.g.][]{Fender2004}. In the case of renewed energy injection episodes, rapid radio brightening can be triggered by rapid energy injection that dominates the energy of the initial blastwave \citep{Laskar2015}. For GRB afterglows with re-flares attributed to energy injection, the optical and X-ray lightcurves usually also show a simultaneous re-brightening \citep{Laskar2015}.  In the case of a TDE, prompt outflows are likely driven by debris interactions during the stellar disruption and circularisation of the debris \citep[e.g.,][]{Lu2020}. Other outflows, by contrast, may be ejected in the form of an accretion induced wind or jet once an accretion disc has formed \citep[e.g.,][]{Ramirez2009,Alexander2016,vanVelzen2011}, and may be delayed from the initial disruption depending on the disc properties and accretion rate \citep{DeColle2012}. At least some TDEs produce highly relativistic jets at early times \citep[e.g., SwiftJ1644; AT2022cmc;][]{Levan2011,Burrows2011,Zauderer2011,Bloom2011,Andreoni2022,Pasham2023}, indicating that TDEs can provide the conditions for the launching of jets. It is currently unclear whether all TDEs produce jets (some less collimated than others) or if the conditions for jet-launching are only reached occasionally in TDEs, although deep observational limits of TDEs suggest powerful jets are rare \citep{vanVelzen2013}. 

State transitions of accretion discs in X-ray binaries are usually associated with changes in radio jet properties \citep{Fender2004}. In X-ray binaries in which stellar mass black holes accrete at varying rates, steady, compact radio-emitting jets are observed during the ``hard" accretion disc state, associated with flat or inverted radio spectra, and X-ray corona emission in hard X-rays \citep[see e.g.][for a review]{Fender2004}. Transient radio-emitting jets are usually launched during accretion disc state transitions during the ``soft" accretion disc state associated with bright, soft X-ray emission and optically thin radio outbursts \citep{Fender2004}. The transient jets associated with the outburst state are assumed to be more relativistic than the steady compact jets in the low state \citep{Fender2004}.%, thought to be due to the collapse of the inner accretion disc channelling more material into the jet. 

It is uncertain how similar TDE accretion disc properties and behaviour are to those of X-ray binaries. Evidence for accretion disc state transitions occurring in TDEs has been observed in a handful of events. \citet{Wevers2021} showed that the X-ray emission from AT2018fyk evolved from a blackbody to power-law dominated spectrum in less than 100\,d. AT2021ehb gradually formed a dominant hard spectral component over 170\,d which softened dramatically within 3\,d while the X-ray flux faded by an order of magnitude \citep{Yao2022}. The pTDE eRASSt J0456-20 also showed a dramatic softening in the X-ray spectrum while fading quickly \citep{Liu2023}. 
At least some TDEs have the conditions required for a hot corona to form, as has been observed in the X-ray spectra of XMMSL1 0740--85 \citep{Saxton2017} and AT2018fyk \citep{Wevers2021}. Additionally, \citet{Hinkle2024} and \citet{Newsome2024} discovered coronal line emission in a number of TDEs, giving tentative evidence of hot coronas in these sources. However, a strong link between X-ray accretion disc emission and radio properties of TDEs has yet to be demonstrated.  

The second radio flare observed from AT2020vwl resulted in an increase in flux density at all frequencies, with the emission peaking first at higher frequencies, similar to the first radio flare. However, the second radio flare rose significantly quicker than the first flare at all frequencies. Through modelling the radio lightcurves, we deduce that the first radio flare contributes very little flux density to the second flare at frequencies $>1$\,GHz (Fig. \ref{fig:lcfits}). 

We investigate whether the
sharp rise of $\nu L_{\nu}\propto t^{4}$ to $t^{7}$ for the second radio flare may be powered by an episode of energy injection. 
The duration of the flare 
%is then
is
dependent on the duration of energy injection, after which the outflow would continue to evolve as a blast wave but with increased kinetic energy \citep{Laskar2015}. 
In this scenario, we begin by noting that the light curves of synchrotron radiation from a non-relativistic shock in the optically thin regime (above the spectral peak in this case, with $\nu_m<\nu_a<\nu<\nu_c$) depend on total shock energy and time as $F_{\nu}\propto E^{0.8}t^{-2.4}$ for $p\approx3$ and $k\approx2.5$ (Appendix~\ref{sec:synappendix}). Assuming the energy rises as $t^m$ during the injection phase, this implies a light curve decline rate of $t^{0.8m-2.4}$. For the highest-frequency 11\,GHz light curve (which is on the optically thin segment throughout and farthest from the spectral peak in our observations), we infer a rise rate of $a_2=7.2^{+2.0}_{-2.8}$ (Table~\ref{tab:lc_fits}), corresponding to an extremely steep energy injection rate of $m=12^{+2.5}_{-3.5}$. This is marginally ($2\sigma$) consistent with the inferred energy increase rate from equipartition arguments, for which fitting the energy growth as a power law with time yields $m_{\rm eq}=4.2\pm1.9$ (Table~\ref{tab:specfits} and Figure~\ref{fig:outflowprops}), making energy injection a possible explanation for the second radio flare. 
\citet{Goodwin2023} found that the first radio flare from AT2020vwl was well-described by a single ejection of energy into an outflow at the onset of the stellar disruption, likely due to collisions between tidal debris streams. 
Energy injection into this outflow could take place if there is slower-moving outflow material travelling behind the main outflow that catches up with the forward shock when the latter decelerates \citep[as seen in $\gamma$-ray bursts, e.g.,][]{sm00,Laskar2015}.    %Whilst it is unlikely the second radio flare was driven by energy injection into the pre-existing outflow, we note that the radio spectral evolution is consistent with an energy injection scenario \citep{Laskar2015}.

Alternatively, the chromatic evolution of the second radio flare, the increase in the peak frequency of the synchrotron spectrum, and the significant increase in energy of the emitting region during the second flare 
%suggest that the second flare may be 
could also be explained if the second flare is 
due to an entirely new outflow launched 160--700\,d after the stellar disruption. 
%
%This begs the question, is late-time energy injection possible for AT2020vwl? 
In Section \ref{sec:discmodelling}, we found evidence for a long-lived accretion disc in the optical and UV emission from the host galaxy of AT2020vwl. Whilst there was no X-ray emission detected from the event at any point in its evolution, the X-ray upper limits allow a strong bound on the accretion rate to be obtained, under the assumption that the late-time X-ray emission is not obscured. We find that the disc is relatively low mass ($\approx0.01\,M_\odot$) and the accretion rate is significantly sub-Eddington ($\sim0.1\,\dot M_{\rm{Edd}}$). 
%It is therefore possible that a radio ejection event could have been driven by the accretion disc in this event at 400-500\,d post-stellar disruption.

%Through modelling the radio lightcurves, we deduce that the first radio flare contributes very little flux density to the second flare at frequencies $<1$\,GHz (Fig. \ref{fig:lcfits}).
In summary, the second outflow carries significantly more energy than the first outflow and is accelerating, potentially indicating a new energy injection (Figure \ref{fig:outflowprops}), strikingly different to the first radio flare observed. %We propose that the second radio flare may be due to a new outflow launched between 170-690\,d after the stellar disruption from the accretion disc. 
The form of this outflow is likely a mildly collimated jet launched at the time of peak accretion rate in the disc (Figure \ref{fig:disc}), similar to transient radio jets that are launched during accretion episodes in X-ray binary systems. An accretion-disc wind is also possible, however less likely as the accretion rate is significantly sub-Eddington, and the second radio flare evolves quickly with increasing energy, different to the accretion rate profile. Energy injection or the launch of a second outflow remain viable models to explain the second flare.

\subsection{Comparison to other TDEs with late-time radio flares}
\begin{figure}
    \centering
    \includegraphics[width=\columnwidth]{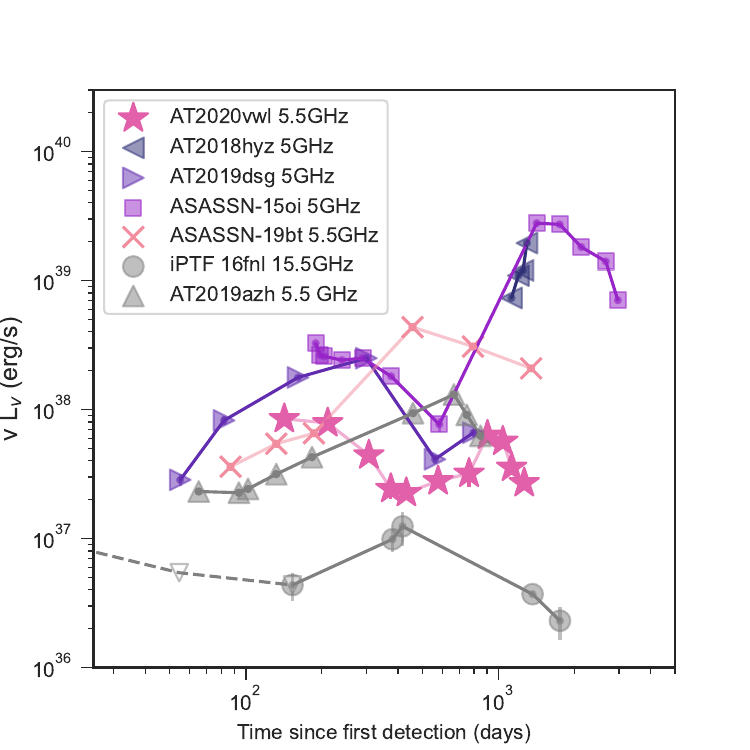}
    \caption{Radio lightcurves of TDEs suggested to have late-rising radio emission or two radio lightcurve components. TDEs shown in grey are consistent with a single outflow ejection and no late-time radio flare. The second radio flare observed from AT2020vwl rises and decays much faster than most other late radio flares observed. Radio data are from \citet[][AT2018hyz]{Cendes2022}, \citet[][ASASSN-19bt]{Christy2024}, \citet[ASASSN-15oi]{Horesh2021,Hajela2024}, \citet[][AT2019dsg]{Cendes2021,Cendes2023}, \citet[][iPTF 16fnl]{Horesh2021b}, and \citet[][AT2019azh]{Goodwin2022}. Empty triangles represent 3$\sigma$ upper limts.}
    \label{fig:TDEcomp}
\end{figure}

Late-time radio flares from TDEs are being discovered with increasing frequency. \citet{Cendes2023} found that in a sample of 23 TDEs, $\approx40\%$ of events had detectable radio emission at late times. Some TDEs with early-time radio emission decay very slowly, over years \citep[e.g., AT2019azh;][]{Goodwin2022}, whilst others decay more rapidly \citep[e.g.,][ASASSN-14li;]{Alexander2016}. Of the TDEs that have shown late-time ($>500$\,d) rising radio emission, only ASASSN-15oi and AT2019dsg have constraining radio observations at early times \citep{Hajela2024,Cendes2021,Stein2021,Horesh2021}. Interestingly, most late-time radio flares observed from TDEs are consistent with a second outflow launched hundreds of days after the stellar disruption \citep{Cendes2023}. A comparison between the radio lightcurve of AT2020vwl and TDEs with well-sampled late-time radio flares is shown in Fig. \ref{fig:TDEcomp}. 

Two distinct radio emission episodes have been discovered in the TDE ASASSN-15oi \citep{Horesh2021,Hajela2024}, and suggested for the TDE ASASSN-19bt \citep{Christy2024}. \citet{Hajela2024} found that for ASASSN-15oi the first radio flare rose as $F_{\nu}\propto t^{4.5}$ and decayed as $F_{\nu}\propto t^{-1}$, whilst the second flare rose with $F_{\nu}\propto t^{4}$ and decayed as $F_{\nu}\propto t^{-1.5}$ until 2660\,d at which time the decay steepened to $F_{\nu}\propto t^{-7}$. The rise and decay rate of the first flare from AT2020vwl are similar (within uncertainties) of the first flare from ASASSN-15oi. AT2020vwl's second flare, however, rose and decayed slightly more quickly than ASASSN-15oi. Similar to AT2020vwl, the first radio flare from ASASSN-15oi is very well-explained by a collision-induced outflow in which material is ejected via stream stream collisions of the stellar debris during the circularisation process \citep{Lu2020,Hajela2024}. As for AT2020vwl, \citet{Hajela2024} found that the outflow that powers the second radio flare seen from ASASSN-15oi carries substantially more energy than the first flare, and they propose that the second radio flare is powered by a second outflow launched around the time of peak accretion at $\approx200$\,d post-optical flare. 
Unlike AT2020vwl, ASASSN-15oi was detected in X-rays, motivating the suggestion that the second radio flare is powered by an accretion-induced wind or mildly collimated jet.

AT2018hyz has shown perhaps the most extreme late-time flare, with the radio luminosity increasing as $L\propto t^5$ at times $>1000$\,d post-optical flare \citep{Cendes2022}. \citet{Cendes2022} modelled the radio spectra of this source and deduced the emission was consistent with a mildly collimated delayed jet that was launched $\approx450$\,d post-optical flare. \citet{Sfaradi2024} proposed that the radio emission from AT2018hyz can be explained by an off-axis prompt relativistic jet launched at the time of the optical flare. Regardless of the mechanism behind the late-rising radio emission from AT2018hyz, the evolution of the radio emission is quite different to the late-time flare from AT2020vwl. The  intrinsic luminosity of the radio emission from AT2018hyz is approximately an order of magnitude higher than that of AT2020vwl and the inferred velocity of the radio-emitting region of the former is significantly higher ($>0.25\,c$ compared to $\approx0.03\,c$). The late radio flare from AT2018hyz showed no sign of beginning to decay at 5\,GHz at 1300\,d post disruption, whereas the second flare of AT2020vwl rose for $\approx500$\,d to peak around 1000\,d post disruption before decaying again. If the late-time radio emission from AT2018hyz and AT2020vwl is explained by the same mechanism, it is clear the outflow in AT2020vwl is significantly slower, less energetic, and shorter-lived than that of AT2018hyz. These differences may be caused by a larger accretion disc in AT2018hyz that is able to launch a stronger jet than that of AT2020vwl. 

\citet{Horesh2021b} reported a potential late-time radio flare from iPTF16fnl, however, due to the intrinsically low luminosity of the radio emission from this source it is consistent with a faint single flare in its radio lightcurve which started around the time of the initial stellar disruption. While, \citet{Sfaradi2022} reported a possible late time flare in the TDE AT2019azh, when taking into consideration the full long-term radio lightcurve presented in \citet{Goodwin2022}, it is evident that this flare is consistent with small amplitude fluctuations in the overall slow rise and decay of the radio emission of this TDE. %Neither of these events show similar radio re-flaring behaviour to AT2020vwl. 

\section{Summary}\label{sec:conclusion}
Our long-term radio observations of the TDE AT2020vwl reveal a second radio flare, beginning at approximately 400\,d post optical flare. The second radio flare rose and decayed quickly at frequencies $>2$\,GHz, unlike late-time flares observed from other TDEs.

We find the second radio flare from AT2020vwl is unlikely to be due to changes in the circumnuclear density distribution or an off-axis jet. We propose that the second radio flare is instead due to either a period of new energy injection into the pre-existing outflow, or a second outflow, launched hundreds of days post stellar disruption, and coincident with the peak accretion rate of the accretion disc that is implied by the optical and UV plateaus in the respective lightcurves. 

There is mounting evidence that TDEs launch multiple types of outflows, with prompt radio-emitting outflows being well-explained by outflows ejected by debris stream collisions, whilst delayed outflows may be launched by the accretion disc which can exist for years after the TDE. ASASSN-15oi and AT2020vwl present the only two TDEs to date that have both prompt and late-type outflows observed in the radio due to continuous radio coverage over many years after discovery. Future observations that track the long-term evolution of radio emission from TDEs and search for any relationship between long-lived accretion discs and delayed radio flares will be crucial in confirming the launching of multiple types of radio outflows in TDEs.

\begin{acknowledgments}
AG is grateful for support from the Forrest Research Foundation. This work was supported by the Australian government through the Australian Research Council’s Discovery Projects funding scheme (DP200102471). This research was supported in part by the National Science Foundation under PHY - 1748958. 
KDA and CC acknowledge support provided by the NSF through award AST-2307668. CC additionally acknowledges NSF and NRAO support from award SOSPA9-007.
This research benefited from interactions that were funded by the Gordon and Betty Moore  Foundation through Grant GBMF5076.
The National Radio Astronomy Observatory is a facility of the National Science Foundation operated under cooperative agreement by Associated Universities, Inc.
We thank the staff of the GMRT that made these observations possible. GMRT is run by the National Centre for Radio Astrophysics of the Tata Institute of Fundamental Research.
The MeerKAT telescope is operated by the South African Radio Astronomy Observatory, which is a facility of the National Research Foundation, an agency of the Department of Science and Innovation.

\end{acknowledgments}

\vspace{5mm}
\facilities{VLA, MeerKAT, GMRT, Swift(XRT and UVOT), ZTF}

% \software{astropy \citep{2013A&A...558A..33A,2018AJ....156..123A},  
%           }

\appendix

\section{Radio Observations}
\begin{longtable}{p{2cm}p{2cm}p{2cm}p{1.5cm}p{2.5cm}}
    \caption{New radio observations of AT2020vwl.}
    \label{tab:radio_obs}\\
 % \begin{tabular}{p{2cm}p{1.5cm}p{1cm}p{2cm}}%p{5cm}}
 %\begin{tabular}
    \hline
    Date (UTC)& Instrument \& configuration & Proposal ID & Frequency (GHz) & Flux Density $\pm$statistical error $\pm$ISS error ($\mu$Jy)\\
\hline
2022-11-01 & VLA-C  & 20B-377 & 1.5  & 186$\pm$89$\pm$74 \\
 & & & 2.77  & 311$\pm$35$\pm$62 \\
 &  && 3.24  & 366$\pm$25$\pm$73 \\
 &  && 4.48  & 216$\pm$39$\pm$17 \\
 &  && 5.51  & 205$\pm$45$\pm$16 \\
 &  && 6.59  & 127$\pm$28$\pm$10 \\
 &  && 7.45  & 82$\pm$32$\pm$2 \\
 &  && 9.1  &82$\pm$11$\pm$2 \\
 &  && 11.12  & 74$\pm$16$\pm$2 \\
 &  && 13.55  & 56$\pm$9$\pm$1 \\
 &  && 16.61  & 29$\pm$9$\pm$1 \\
 \hline
2023-02-01 & GMRT & 43$\_$046 & 0.65 & 292$\pm$42$\pm$58 \\
2023-02-01 & GMRT & 43$\_$046 & 1.26 & 417$\pm$41$\pm$167 \\
 \hline
2023-04-22 & GMRT & 44$\_$022 & 0.65  & 156$\pm$54$\pm$31 \\
2023-04-01 & VLA-B & 23A-202 & 1.263  & 604$\pm$53$\pm$242 \\
 &  && 1.65   & 471$\pm$95$\pm$188 \\
 &  && 1.9  &742$\pm$43$\pm$297 \\
 &  && 2.24   &699$\pm$44$\pm$140 \\
 &  && 2.75   &639$\pm$21$\pm$128 \\
 &  && 3.25  & 578$\pm$18$\pm$116 \\
 &  && 3.75   &562$\pm$17$\pm$112 \\
 &  && 4.49   &488$\pm$23$\pm$39 \\
 &  && 5.51   &404$\pm$22$\pm$32 \\
 &  && 6.49   &414$\pm$24$\pm$33 \\
 &  && 9.0   &325$\pm$14$\pm$7 \\
 &  && 11.0  & 274$\pm$13$\pm$6 \\
 \hline
2023-08-31 & GMRT &44$\_$022 &  0.65 & 689$\pm$59$\pm$138 \\
2023-08-31 & GMRT &44$\_$022 &  1.26 & 601$\pm$64$\pm$240 \\
2023-08-01 & VLA-A& 23A-202 & 1.14  & 692$\pm$148$\pm$277 \\
 & & & 1.4  & 639$\pm$39$\pm$256 \\
 & & & 1.65  & 598$\pm$71$\pm$239 \\
 & & & 1.9  &590$\pm$31$\pm$236 \\
 & & & 2.24  & 696$\pm$43$\pm$139 \\
 & & & 2.75   &689$\pm$26$\pm$138 \\
 & & & 3.25   &625$\pm$21$\pm$125 \\
 & & & 3.75   &549$\pm$19$\pm$110 \\
 & & & 4.49   &512$\pm$26$\pm$41 \\
 & & & 5.51  &363$\pm$28$\pm$29 \\
 & & & 6.49   &338$\pm$28$\pm$27 \\
 & & & 7.51   &283$\pm$26$\pm$6 \\
 & & & 9.0  &188$\pm$16$\pm$4 \\
  & & & 11.0  &189$\pm$37$\pm$4 \\
 \hline
2023-11-15 & GMRT & 45$\_$123  & 0.65  & 595$\pm$93$\pm$119 \\ 
%2023-11-15 & GMRT  & 45$\_$123 & 1.26  & \textcolor{red}{to fill in}\\ 
2023-11-10 & VLA-D & 23B-078  & 2.75  & 425$\pm$39$\pm$85 \\
 &  && 3.24  & 434$\pm$32$\pm$87 \\
 &  && 3.75  & 441$\pm$103$\pm$88 \\
 &  && 4.49  & 298$\pm$41$\pm$24 \\
 &  && 5.51  & 228$\pm$22$\pm$18 \\
 &  && 6.49  & 210$\pm$25$\pm$17 \\
 &  && 7.51  & 180$\pm$24$\pm$4 \\
 &  && 9.0   &133$\pm$17$\pm$3 \\
 &  && 11.0  & 89$\pm$21$\pm$2 \\
 \hline
 2024-02-27 & MeerKAT & SCI-20230907-AG-01 & 0.82  & 1131$\pm$45$\pm$226 \\
 2024-03-26 & VLA-C & 24A-159 & 2.24  & 568$\pm$77$\pm$114 \\
 &  && 2.75  & 366$\pm$35$\pm$73 \\
 &  && 4.5   &201$\pm$26$\pm$16 \\
 &  && 5.5   &173$\pm$21$\pm$14 \\
 & & & 6.49  & 178$\pm$19$\pm$14 \\
 & & & 7.51  & 111$\pm$18$\pm$2 \\
 & & & 9.0  &106$\pm$14$\pm$2 \\
 & & & 11.0  & 79$\pm$16$\pm$2 \\
 & & & 13.5  & 60$\pm$9$\pm$1 \\
 & & & 16.62 & 47$\pm$10$\pm$1 \\
 \hline
\end{longtable}

\section{Spectral fitting parameters and outflow constraints}

\begin{sidewaystable*}[]
    \centering
        \caption{Modelled radio spectra and outflow properties for AT2020vwl. The observations between 97 and 532 days post-optical flare were published in \citet{Goodwin2023}.}
    \label{tab:specfits}
    \begin{tabular}{llllllllllllll}
\hline
Date (UTC) & $\delta$t (d)$^{a}$ & $F_{\rm{p}}$ (mJy) & $\nu_{\rm{m}}$ (GHz) & $\nu_{\rm{p}}$ (GHz) & $p$ & log $E$ (erg) &  log $R$ (cm) &
 $\beta$ (c) & log $n_{e}$ (cm$^{-3}$) & log $B$ (G) & log $M_{\rm ej}$ (M$_{\odot}$) \\

\hline
Spherical geometry & &&&&&&&&&&&& \\
\hline
27/2/2021 & 142 &
0.59$\pm$0.04& 4.02$\pm$1.09 & 7.44$\pm$0.63 & 2.91$\pm$    0.27 &
48.07$\pm$0.07 & 16.11$\pm$0.06 &    0.035$\pm$0.005 &     3.80$\pm$0.55 &     -0.14$\pm$0.41 &     -2.98$\pm$0.10\\
7/5/2021 & 211 &
0.53$\pm$0.04& 3.06$\pm$0.90 & 5.74$\pm$0.66 & 2.92$\pm$    0.26 &
48.13$\pm$0.08 & 16.21$\pm$0.07 &    0.029$\pm$0.005 &     3.59$\pm$0.59 &     -0.24$\pm$0.45 &     -2.76$\pm$0.11\\
11/8/2021 & 307 &
0.37$\pm$0.07& 1.47$\pm$0.52 & 2.75$\pm$0.47 & 3.02$\pm$    0.22 &
48.35$\pm$0.13 & 16.46$\pm$0.09 &    0.036$\pm$0.008 &     3.05$\pm$0.81 &     -0.52$\pm$0.58 &     -2.72$\pm$0.16\\
18/10/2021 & 375 &
0.23$\pm$0.14& 1.19$\pm$0.59 & 2.25$\pm$0.65 & 3.05$\pm$    0.22 &
48.21$\pm$0.35 & 16.45$\pm$0.17 &    0.029$\pm$0.012 &     2.94$\pm$1.57 &     -0.57$\pm$1.04 &     -2.67$\pm$0.39\\
14/12/2021 & 432 &
0.14$\pm$0.04& 1.43$\pm$0.72 & 2.75$\pm$0.84 & 2.97$\pm$    0.26 &
47.79$\pm$0.20 & 16.26$\pm$0.13 &    0.016$\pm$0.005 &     3.11$\pm$1.13 &     -0.49$\pm$0.82 &     -2.59$\pm$0.24\\
8/5/2022 & 577 &
0.27$\pm$0.09& 0.81$\pm$0.36 & 1.55$\pm$0.41 & 3.08$\pm$    0.21 &
48.48$\pm$0.20 & 16.65$\pm$0.12 &    0.030$\pm$0.008 &     2.62$\pm$1.10 &     -0.73$\pm$0.76 &     -2.42$\pm$0.24\\
1/11/2022 & 765 &
0.16$\pm$0.07& - & 2.20$\pm$0.66 & 3.09$\pm$    0.21 &
48.05$\pm$0.27 & 16.39$\pm$0.15 &    0.013$\pm$0.004 &     2.98$\pm$1.39 &     -0.55$\pm$0.94 &     -2.10$\pm$0.31\\
1/4/2023 & 905 &
0.48$\pm$0.06& - & 3.55$\pm$0.40 & 2.90$\pm$    0.26 &
48.27$\pm$0.12 & 16.39$\pm$0.10 &    0.011$\pm$0.003 &     3.17$\pm$0.87 &     -0.45$\pm$0.66 &     -1.75$\pm$0.16\\
1/8/2023 & 1041 &
0.56$\pm$0.08& - & 2.05$\pm$0.27 & 3.09$\pm$    0.20 &
48.75$\pm$0.13 & 16.68$\pm$0.11 &    0.018$\pm$0.004 &     2.80$\pm$0.95 &     -0.64$\pm$0.69 &     -1.71$\pm$0.17\\
10/11/2023 & 1128 &
0.31$\pm$0.08& - & 2.15$\pm$0.53 & 3.04$\pm$    0.23 &
48.37$\pm$0.18 & 16.53$\pm$0.13 &    0.012$\pm$0.003 &     2.87$\pm$1.11 &     -0.61$\pm$0.80 &     -1.73$\pm$0.22\\
26/3/2024 & 1265 &
0.53$\pm$0.23& - & 0.85$\pm$0.31 & 3.11$\pm$    0.18 &
49.12$\pm$0.26 & 17.05$\pm$0.16 &    0.034$\pm$0.012 &     2.06$\pm$1.41 &     -1.02$\pm$0.97 &     -1.90$\pm$0.30\\
\hline
Conical geometry & &&&&&&&&&&&& \\
\hline
& 142 && &  &  & 48.59$\pm$0.07 & 16.49$\pm$0.06 &    0.080$\pm$0.012 &     3.27$\pm$0.55 &     -0.41$\pm$0.41 &     -3.16$\pm$0.10\\
& 211 && &  &  & 48.66$\pm$0.08 & 16.58$\pm$0.07 &    0.067$\pm$0.011 &     3.06$\pm$0.59 &     -0.51$\pm$0.45 &     -2.95$\pm$0.11\\
& 307 & & &  &  & 48.89$\pm$0.13 & 16.84$\pm$0.09 &    0.082$\pm$0.017 &     2.52$\pm$0.81 &     -0.78$\pm$0.58 &     -2.89$\pm$0.16\\
& 375 & & &  &  & 48.75$\pm$0.35 & 16.83$\pm$0.17 &    0.067$\pm$0.026 &     2.41$\pm$1.57 &     -0.84$\pm$1.04 &     -2.85$\pm$0.39\\
& 432 & & &  &  & 48.32$\pm$0.20 & 16.63$\pm$0.13 &    0.038$\pm$0.011 &     2.57$\pm$1.13 &     -0.75$\pm$0.82 &     -2.79$\pm$0.24\\
& 577 & & &  &  & 49.02$\pm$0.20 & 17.03$\pm$0.12 &    0.068$\pm$0.019 &     2.09$\pm$1.10 &     -1.00$\pm$0.76 &     -2.60$\pm$0.24\\
& 765 & & &  &  & 48.59$\pm$0.27 & 16.76$\pm$0.15 &    0.029$\pm$0.010 &     2.45$\pm$1.39 &     -0.82$\pm$0.94 &     -2.30$\pm$0.31\\
& 905 & & &  &  & 48.80$\pm$0.12 & 16.77$\pm$0.10 &    0.025$\pm$0.006 &     2.63$\pm$0.87 &     -0.72$\pm$0.66 &     -1.96$\pm$0.16\\
& 1041 & &  &  & & 49.29$\pm$0.13 & 17.05$\pm$0.11 &    0.042$\pm$0.010 &     2.27$\pm$0.95 &     -0.91$\pm$0.69 &     -1.90$\pm$0.17\\
& 1128 & & &  &  & 48.91$\pm$0.18 & 16.91$\pm$0.13 &    0.028$\pm$0.008 &     2.33$\pm$1.11 &     -0.87$\pm$0.80 &     -1.93$\pm$0.22\\
& 1265 & & &  &  & 49.66$\pm$0.26 & 17.43$\pm$0.16 &    0.078$\pm$0.028 &     1.53$\pm$1.41 &     -1.28$\pm$0.97 &     -2.08$\pm$0.30\\
    \end{tabular}
    $^a$ $\delta t$ is measured with respect to the beginning of the optical flare, MJD 59130. 
\end{sidewaystable*}

\section{Optically thin synchrotron emission from non-relativistic outflows in stratified media}
\label{sec:synappendix}
Consider a point explosion of energy, $E$, driving a non-relativistic shock with speed $\beta_{\rm sh} = \dot{r}_{\rm sh}/c$ into a stratified medium with density structure, $\rho\propto r^{-k}$. The shock speed and radius as a function of time are given by the Sedov-Taylor-von Neumann solution, with $r_{\rm sh}\propto E^{\frac{1}{5-k}}t^{\frac{2}{5-k}}$ and $\beta_{\rm sh}\propto E^{\frac{1}{5-k}}t^{-\frac{3}{5-k}}$, such that $\rho r^3\beta_{\rm sh}^2={\rm const}$. We assume that a fixed fraction of the internal energy density of the post-shock fluid is given to magnetic fields ($B^2\propto \epsilon_{\rm B}\rho \beta_{\rm sh}^2$), such that 
\begin{equation}
B\propto \rho^{\frac12}\beta_{\rm sh} \propto E^{\frac{2-k}{2(5-k)}}t^{-\frac{3}{5-k}}.
\end{equation}
The characteristic synchrotron frequency for electrons with Lorentz factor, $\gamma_e$ moving at angle, $\alpha$ in a magnetic field, $B$ (in SI units) is given by 
\begin{equation}
\nu\sim\frac{1}{1+z}\frac{3}{4\pi}\gamma_e^2\frac{eB}{m_e}\sin{\alpha}\propto\gamma_e^2B. 
\end{equation}
This implies that the Lorentz factor of electrons that dominate the emission at an observing frequency, $\nu$ is given by 
\begin{equation}
    \gamma_e \propto \nu^{1/2}B^{-1/2}\propto \nu^{1/2}E^{-\frac{2-k}{4(5-k)}}t^{\frac{3}{2(5-k)}}. \label{eq:gammaofnu}
\end{equation}
Since the radiation emitted by these electrons peaks at $P_\nu$ such that $\nu P_\nu\sim P_{\rm syn}\propto \gamma_e^2 B^2$ is the synchrotron energy loss rate, the specific power, 
\begin{equation}
    P_{\nu} \propto \frac{P_{\rm syn}}{\nu} \propto B \propto E^{\frac{2-k}{2(5-k)}}t^{-\frac{3}{5-k}} \label{eq:Pnu}
\end{equation}
is independent of $\gamma_e$. 

We assume that the shock accelerates electrons into a power law distribution in electron Lorentz factor, $\gamma_e$, such that $dN_e/d\gamma_e\propto\gamma^{-p}$ above a minimum Lorentz factor, $\gamma_m$. 
The standard assumption that electrons at $\gamma_e>\gamma_m$ carry a fraction $\epsilon_e$ of the shock energy leads to $\gamma_m \approx \epsilon_e \frac{\beta_{\rm sh}^2}{2} \frac{m_p}{m_e}$ for non-relativistic shocks. For values of $\epsilon_e\lesssim0.1$, this expression breaks down when $\beta_{\rm sh}\lesssim0.1$. In this regime, it is important to include the deviation of the electron participation fraction from unity \citep{ew05,sg13}. Following \citep{sg13}, we have 
\begin{equation}
   \gamma_m = \max\left[1,x\right], f_{\rm NT} = \min\left[1,x,x^{{p-1}/2}\right], x\equiv\epsilon_e\frac{\beta_{\rm sh}^2}{2} \frac{m_p}{m_e},
\end{equation}
where $f_{\rm NT}$ is the fraction of particles accelerated into the non-thermal distribution. For $\epsilon_e\beta_{\rm sh}^2\lesssim 2m_e/m_p\approx\times10^{-3}$, we have $x<1$, $\gamma_m\sim1$, and $f_{\rm NT}=\min\left[x,x^{(p-1)/2}\right]$. This leads to the following choices for $f_{\rm NT}$ 
\begin{equation}
  f_{\rm NT}=
  \begin{cases}
    x^{(p-1)/2}\propto\beta_{\rm sh}^{p-1}, & p\gtrsim3 \\    
    x\propto \beta_{\rm sh}^2, & 2<p<3
  \end{cases}    
  \label{eq:fnt}
\end{equation}

The specific luminosity above the synchrotron self-absorption break is then given by the product of the number of electrons emitting near frequency $\nu$ and the specific power per electron,
\begin{equation}
    L_\nu = N_e{\nu}P_\nu \propto f_{\rm NT}N_e\left(\gamma_e/\gamma_m\right)^{1-p}P_\nu, \label{eq:Lnu}
\end{equation}
where
\begin{equation}
    N_e \propto \rho r_{\rm sh}^3 \propto r^{3-k}
\end{equation}
is the total number of electrons swept up by the shock. Taking $\gamma_m\sim1$ and substituting the expression for $f_{\rm NT}$ from equation~\ref{eq:fnt}, along with $\gamma_e$ from equation~\ref{eq:gammaofnu}, and $P_\nu$ from equation~\ref{eq:Pnu} into equation~\ref{eq:Lnu}, the observed luminosity in the optically thin regime is given by
\begin{equation}
    L_\nu\propto 
    \begin{cases}
        \nu^{\frac{1-p}{2}}E^{\frac{(p-1)(6-k)+(16-6k)}{4(5-k)}} t^{-\frac{2(3-k)(p-3)+3(p+1)}{2(5-k)}}, & p\gtrsim3 \\
        \nu^{\frac{1-p}{2}}E^{\frac{6(4-k)+(k-2)(1-p)}{4(5-k)}} t^{-\frac{3(p+1)}{2(5-k)}}, & 2<p<3.
    \end{cases}
\end{equation}
These give the consistent solution, $L_\nu\propto E^{\frac{7-2k}{5-k}}t^{-\frac{6}{5-k}}$ for $p=3$. In addition for the typical values of $k\approx2.5$ inferred for radio TDEs, this suggests that the optically thin synchrotron spectrum, $L_\nu\propto E^{0.8}t^{-2.4}$ for $p\approx3$. 

\bibliography{bibfile}{}
\bibliographystyle{aasjournal}



\end{document}